%% file: apssamp.tex
\documentclass[%
 reprint,
superscriptaddress,
 amsmath,amssymb,
 aps,
]{revtex4-2}
\usepackage{calc}
\usepackage{pgf}
\usepackage{graphicx}
\usepackage{dcolumn}
\usepackage{bm}
\usepackage{xcolor}
\usepackage{float}
\usepackage{siunitx}
\usepackage{subfigure}
\usepackage{siunitx}
\usepackage{pgf}
\DeclareSIUnit \cycle {cyc}
\DeclareSIUnit{\belmilliwatt}{Bm}
\DeclareSIUnit{\dBm}{\deci\belmilliwatt}
\usepackage{glossaries}
\usepackage{lineno}
\usepackage{etoolbox}

\newtoggle{record}
\newcounter{highlights}
\newcommand{\highlightslist}{}

\makeatletter
\newcommand{\highlightsentry}[2]{%
  \g@addto@macro\highlightslist{\par Line #1: #2}%
}

\newcommand{\addhighlight}[1]{%
  \ifnum\value{linenumber}>0
    \stepcounter{highlights}%
    \edef\temp{\thelinenumber}%
    \expandafter\highlightsentry\expandafter{\temp}{#1}%
  \fi
}
\makeatother

\newcommand{\review}[1]{\textcolor{black}{#1}\iftoggle{record}{\addhighlight{#1}}{}}

\newcommand{\reviewCaption}[1]{%
  \textcolor{black}{#1}%
}

\newcommand{\stoprecording}{\togglefalse{record}}
\begin{document}

\graphicspath{{Figs/}}
\preprint{APS/123-QED}

\title{Free Space Optical Frequency Comparison Over Rapidly Moving Links}

\author{Shawn M. P. McSorley}
\affiliation{International Centre for Radio Astronomy Research, The University of Western Australia, Crawley, WA 6009, Australia}
\affiliation{These two authors contributed equally}
\author{Benjamin P. Dix-Matthews}
\affiliation{International Centre for Radio Astronomy Research, The University of Western Australia, Crawley, WA 6009, Australia}
\affiliation{These two authors contributed equally}
\author{Alex M. Frost}
\affiliation{International Centre for Radio Astronomy Research, The University of Western Australia, Crawley, WA 6009, Australia}
\author{Ayden S. McCann}
\affiliation{International Centre for Radio Astronomy Research, The University of Western Australia, Crawley, WA 6009, Australia}
\author{Skevos F. E. Karpathakis}
\affiliation{International Centre for Radio Astronomy Research, The University of Western Australia, Crawley, WA 6009, Australia}
\author{David R. Gozzard}
\affiliation{International Centre for Radio Astronomy Research, The University of Western Australia, Crawley, WA 6009, Australia}
\author{Shane M. Walsh}
\affiliation{International Centre for Radio Astronomy Research, The University of Western Australia, Crawley, WA 6009, Australia}
\author{Sascha W. Schediwy}
\affiliation{International Centre for Radio Astronomy Research, The University of Western Australia, Crawley, WA 6009, Australia}


\begin{abstract}
The comparison of optical reference frequency signals over free-space optical links is limited by the relative motion between local and remote sites. \review{For ground to low earth orbit comparison, the expected Doppler shift and Doppler rate typically reach $\pm$\qty{4}{\giga\hertz} at \qty{100}{\mega\hertz\per\second}, which prevents the narrow-band detection required to compare optical frequencies at the highest levels of stability.} We demonstrate a system capable of optical frequency comparison in the presence of \review{these significant} Doppler shifts, using an electro-optic phase modulator with an actuation bandwidth of \qty{10}{\giga\hertz}, \review{which will} enable ground-to-space frequency comparison. This system was demonstrated over a retro-reflected drone link, with a maximum line-of-sight velocity of \qty{15}{\metre\per\second} and Doppler shift of \review{\qty{19}{\mega\hertz} at a Doppler rate of \qty{1}{\mega\hertz\per\second}}. \review{The best fractional frequency stability obtained was \num{7d-18} at an integration time of \qty{5}{\second}}. \review{These results are an important step toward ground to low earth orbit optical frequency comparison, providing a scalable terrestrial testbed.}
\end{abstract}

\maketitle

\review{The comparison of ultra-stable optical clocks from ground-to-space} will allow for unprecedented tests of fundamental physics. Optical clocks are already capable of testing Einstein's theory of General Relativity in the weak field regime \cite{Takamoto2020, Takamoto2022}, providing extremely precise measurements of gravitational redshift. Their impact will be seen in geodesy, providing new means of defining and measuring the geoid \cite{LionG2017, Takamoto2022}, \review{while also} contributing to the redefinition of the SI second \cite{Lodewyck2019}.

The fractional frequency stability achieved by current state-of-the-art optical clocks is on the order of \num{d-18} \cite{Takamoto2022}. The comparison of these ultra-stable optical reference frequency signals (herein referred to as optical references) is not possible with conventional radio-frequency techniques, which are limited by their low carrier frequency. The past 20 years has seen rapid development on a variety of optical technologies capable of transferring or comparing optical references \cite{DROSTE2015524, Schediwy2013, Gozzard2021, KangHyunJay2019Ftoc, YANG2022, Calosso14,Bercy14,Bergeron2019,caldwell2023quantum, DixMatthews2021, Chiodo2013, McSorley2023, Dix-Matthews2023-TOF}.

\review{Two-way frequency transfer (TWFT) can be used to compare optical frequencies in post processing \cite{Calosso14,Bercy14}, similar to two-way time transfer techniques demonstrated in \cite{Bergeron2019} and \cite{caldwell2023quantum}. This technique is limited by the detectors used to measure the optical reference phase at both the local and remote site. A benefit of TWFT is it requires less launch power than similar reflected Doppler cancellation techniques \cite{Gozzard2021, Dix-Matthews2023-TOF, McSorley2023}, which allows for smaller terminal size, weight and power requirements.}

\review{To ensure sufficient signal-to-noise, the link loss necessitates that the two-way technique have narrow bandwidth photodetectors (PDs) and phase detectors. These bandwidths are typically less than \qty{100}{\mega\hertz} and \qty{10}{\kilo\hertz} respectively, for extreme attenuation \cite{Chiodo2013, DICK2008}. Optical frequency comb TWFT techniques have been demonstrated with Doppler shifts up to \qty{20}{\mega\hertz}  \cite{Bergeron2019}.}

\review{This technique is limited by the large Doppler shift and Doppler rates of the optical carrier experienced over ground-to-space \cite{Dix-Matthews2020, DixMatthews2023-EDV, Dix-Matthews2023-TOF}. For example, optical TWFT to  satellites in low earth orbit (LEO), operating at a wavelength of \qty{1550}{\nano\metre}, are expected to experience Doppler shifts on the order of $\pm \qty{4}{\giga\hertz}$ at rates of \qty{100}{\mega\hertz\per\second} \cite{Shen21}. This will rapidly shift the optical frequency outside of the narrow detection bandwidth. To our knowledge, no technique has been put forward capable of being scaled to the Doppler shift and Doppler rates experienced in ground-to-space clock comparison.}

\begin{figure*}[ht!]
    \centering
    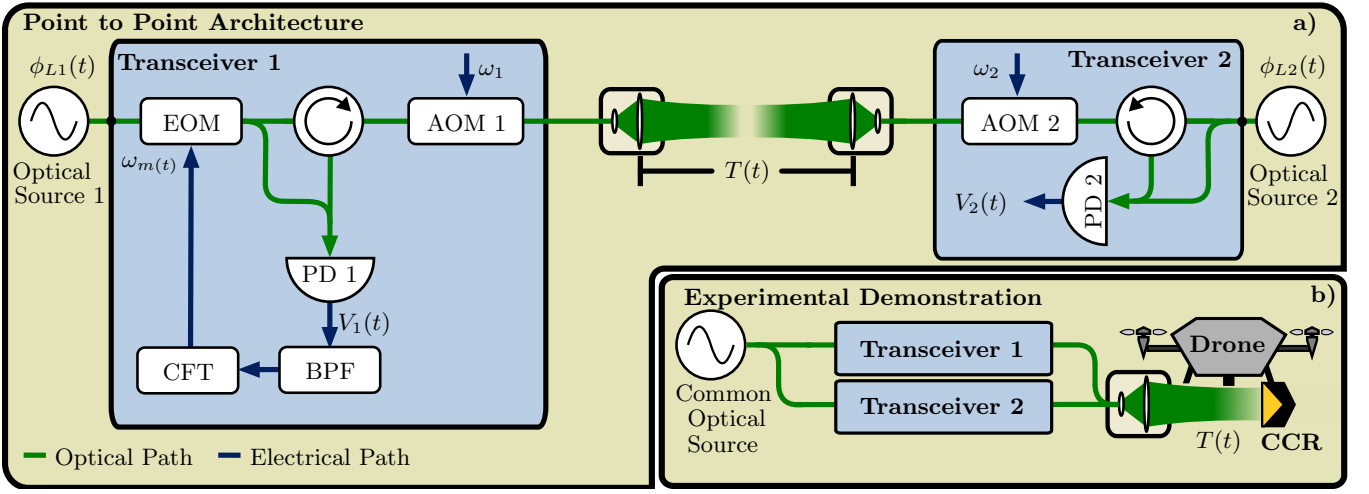
    \caption{
    \label{fig:ExpDiagram1}\reviewCaption{Point-to-point system diagram and experimental demonstration diagram for free-space optical frequency transfer to rapidly moving targets. EOM, electro-optic modulator; AOM, acousto-optic modulator; PD, photodetector; BPF, band-pass filter; CFT, coarse frequency tracking; CCR, corner-cube retro-reflector; $\omega_m(t)$, EOM drive frequency; $\omega_1$ and $\omega_2$, AOM 1 and AOM 2 drive frequency; $V_1(t)$ and $V_2(t)$, PD1 and PD2 voltages used to obtain $\phi_1$ and $\phi_2(t)$; $T(t)$, optical time-of-flight. (a) Point-to-Point architecture. (b) folded experimental demonstration.}}
\end{figure*}

In this Letter, we describe a system for optical frequency comparison over rapidly moving FSO links, using TWFT. The system, shown in Fig.~\ref{fig:ExpDiagram1}(a), is capable of compensating Doppler shifts up to $\pm \qty{10}{\giga\hertz}$. We demonstrate this system over a moving retro-reflected link to an airborne drone with a maximum line-of-sight velocity of \qty{15}{\metre\per\second} \review{at an acceleration of \qty{0.9}{\metre\per\second\squared}}, limited by the maximum speed of the drone, or an equivalent Doppler shift of \qty{18}{\mega\hertz} \review{at a Doppler rate of \qty{1}{\mega\hertz\per\second}}. This exceeds the Doppler shift of similar \review{continuous-wave} demonstrations by an order of magnitude \cite{Dix-Matthews2023-TOF, DixMatthews2023-EDV}, \review{and can be scaled to a LEO link.} \review{A fractional frequency stability of \num{7d-18} at an integration time of \qty{5}{\second} was obtained.}

This system comprises two \review{\qty{1550}{\nano\metre}} optical transceivers, each containing an optical reference. The two optical references can be modelled in terms of their phase denoted by $\phi_{L1}(t)=\omega_{L1}t+\Delta\phi_{L1}(t)$ and $\phi_{L2}(t)=\omega_{L2}t+\Delta\phi_{L2}(t)$. Here, $\omega_{L1}$  and $\omega_{L2}$ are the nominal frequency of each optical reference. Phase fluctuations around these nominal frequencies are denoted by $\Delta\phi_{L1}$ and $\Delta\phi_{L2}$. \review{The frequency difference, $\Delta\dot{\phi}_L(t)=\Delta\dot\phi_{L2}(t) - \Delta\dot\phi_{L1}(t)$, is the primary metric of interest for frequency comparison.}

Independent comparisons of $\phi_{L1}(t)$ and $\phi_{L2}(t)$ are made at both transceiver \review{photodetectors (PDs)} using \review{phase locked loops (PLLs)} to determine $\Delta\phi_L(t)$. The comparisons are made by forming two optical interferometers which provide two radio-frequency (RF) beat notes that encode $\Delta\phi_L(t)$, corrupted by the free-space link noise. The RF beat notes are centred on a known frequency using two acousto-optic modulators (AOM), with frequency shifts denoted by $\omega_1$ and $\omega_2$.

For optical frequency comparison over rapidly moving links, where the optical signal experiences Doppler shifts exceeding the PD bandwidth, it becomes impossible to infer $\Delta\phi_L(t)$ without further system modifications. Furthermore, Doppler correction using AOMs is not feasible for frequency shifts exceeding the AOM actuation bandwidth (typically on the order of only \qty{1}{\mega\hertz}).

To bypass these limitations, Doppler correction is achieved using an electro-optic modulator (EOM), which typically has a much greater actuation bandwidth (up to \qty{100}{\giga\hertz}). In this system, a phase modulating EOM is driven at a frequency $\omega_m(t)$, which creates two optical sidebands at frequencies \review{$\omega_{L1}-\omega_m(t)$ and $\omega_{L1}+\omega_m(t)$. For convenience, the EOM phase, $\phi_m(t)=\int^t_0 \omega_m(\tau)d\tau$, is separated into a static frequency term, $\omega_m$, and a time varying term, $\Delta\phi_m(t)$, such that $\phi_m(t) = \omega_m t + \Delta\phi_m(t)$.}

\review{The instantaneous phase at both PDs is then given by,
\begin{align}
 \phi_{1}(t)&=\phi_{L2}(t-T(t))-\phi_{L1}(t) +  \phi_{\text{m}}(t) \\ &+\omega_1t +\omega_2\{t-T(t)\}\nonumber \\
 &= (\omega_1+\omega_2+\omega_m+\omega_{L2}-\omega_{L1})t\nonumber\\
 &-\dot{T}(t)\{\omega_{L2}+\omega_2\} + \Delta\phi_L(t) + \Delta\phi_m(t)\nonumber,\text{ and}\\
 \phi_{2}(t)&=\phi_{L1}(t-T(t))-\phi_{L2}(t) + \phi_{\text{m}}(t-T(t)) \\ &+\omega_2t +\omega_1\{t-T(t)\}\nonumber\\
 &= (\omega_1+\omega_2+\omega_m+\omega_{L1}-\omega_{L2})t\nonumber\\
 -\dot{T}&(t)\{\omega_{L1}+\omega_1 + \omega_m(t)\} - \Delta\phi_L(t) + \Delta\phi_m(t-T(t)),\nonumber
\end{align}
where $T(t)$ is the optical time-of-flight of the link. The symmetric optical sidebands created by the EOM ensure that $\phi_{m}(t)$ is added to both PD phases. As the transceiver laser sources are expected to be locked to optical atomic clocks, their large-scale frequency differences will be slowly varying. Thus, it is assumed that $\Delta\phi_{L1,L2}(t-T(t))\approx\Delta\phi_{L1,L2}(t)$.
}

\review{Assuming the nominal frequencies $\omega_{L1}$ and $\omega_{L2}$ are equal, the beat frequency of PD1 and PD2 are both centered on $\omega_{beat}=\omega_1+\omega_2+\omega_m$, with Doppler shift terms proportional to $\dot{T}(t)(\omega_{L1,L2})$. Coarse frequency control of the EOM can be used to keep the total instantaneous frequency within the detectable bandwidth of each PD; this can be achieved with either a servo loop, or an \textit{a priori} sweep.}


\newlength{\myheight}
\newlength{\mywidth}
\settowidth{\mywidth}{\includegraphics{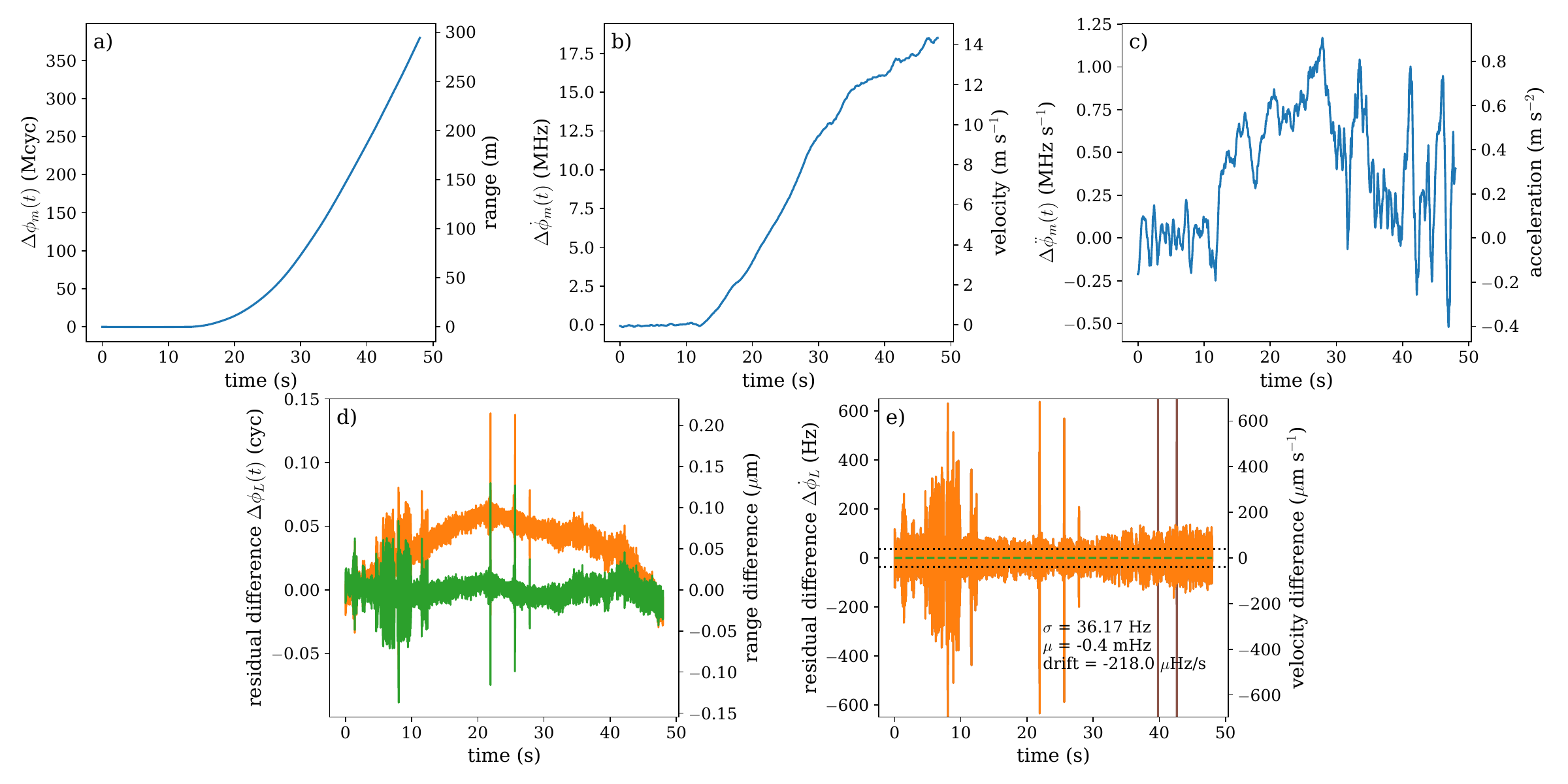}}
\settoheight{\myheight}{\includegraphics{highspeed_timeseries.pdf}}
\pgfmathsetmacro{\aspectratio}{\mywidth/\myheight}
\pgfmathsetmacro{\tswords}{300 / (0.5 * \aspectratio) + 40}

\begin{figure*}[bt!]
    \centering
    \includegraphics[width=\textwidth]{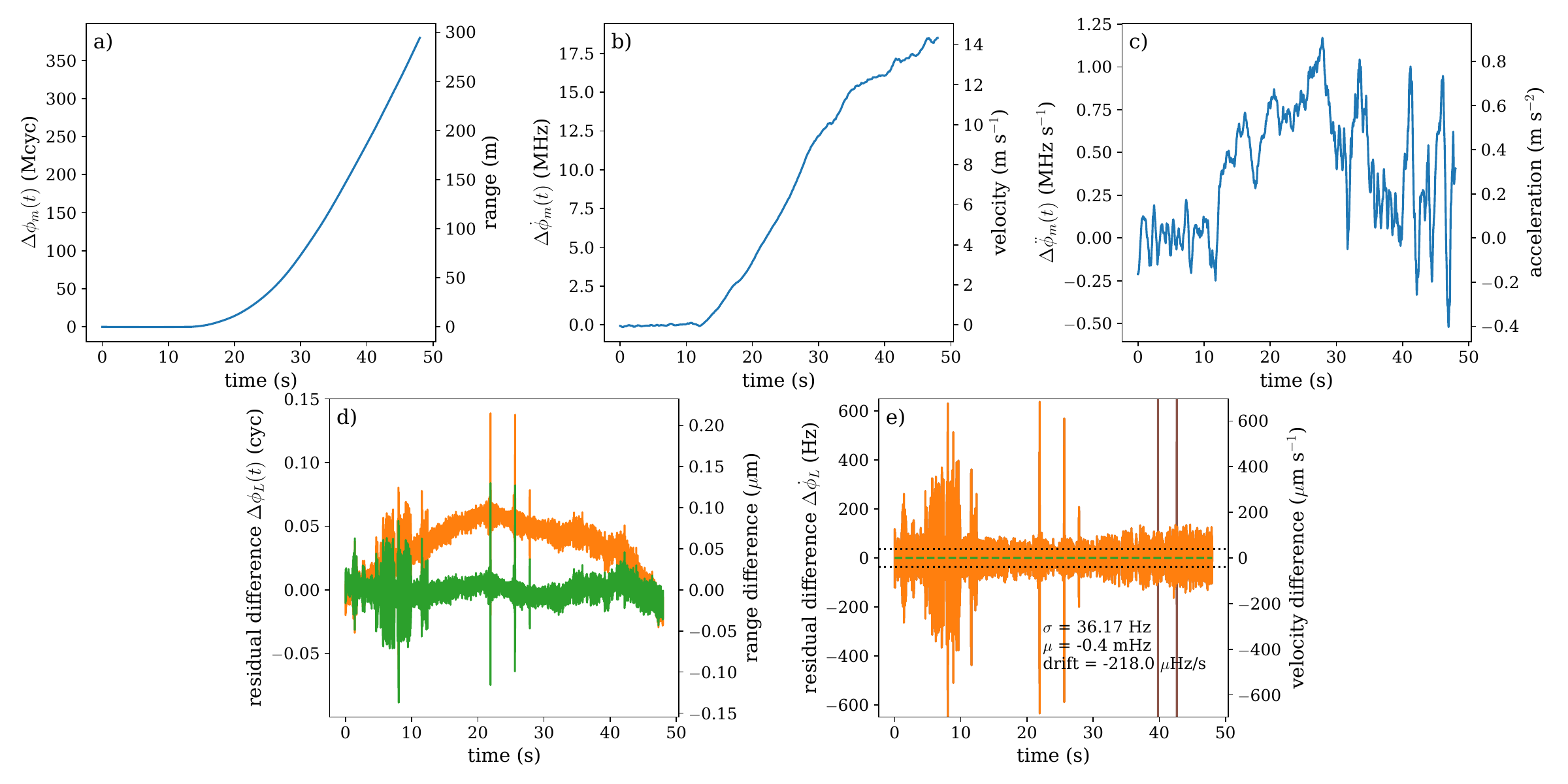}
    \caption{\reviewCaption{Time series obtained for a drone pass with maximum speed of \qty{15}{\metre\per\second} and Doppler shift of \qty{20}{\mega\hertz}. Shown on the left-hand y-axes: (a) the accumulated phase of the tracking EOM; (b) the instantaneous frequency of the tracking EOM; (c) the time derivative of the tracking frequency; (d) the residual phase difference from Equation \ref{eq:noiseqn} (orange), and the residual phase difference with frequency drift removed (green); (e) the instantaneous residual frequency difference (orange), masked cycle slips (brown), 1-sigma error bars (dotted), frequency drift obtained from (d) (dashed green) and the mean frequency bias. The frequency drift in plot (d) is determined with a parabolic fit. In plot (e), frequency impulses exceeding \qty{800}{\hertz}, shown in brown, are masked. Shown on the right-hand y-axes are the associated time-of-flight parameters, with phase corresponding to range, frequency corresponding to velocity, and the time derivative of frequency corresponding to acceleration.} }
    \label{fig:timeseries}
\end{figure*}

\review{By setting the initial PLL frequency to $\omega_{beat}$, two phase measurements $\Delta\phi_2(t)=\phi_2(t)-\omega_{beat}t$ and $\Delta\phi_1(t)=\phi_1(t)-\omega_{beat}t$ are obtained. The difference $\Delta\phi_L(t)$ can then be determined as,
\begin{align}
    \Delta\phi_L(t)&=-\frac{k_2}{k_1+k_2}\Delta\phi_{\mathrm{1}}(t)+\frac{k_1}{k_1+k_2}\Delta\phi_{\mathrm{2}}(t)\\
    &+ \frac{k_2}{k_1+k_2}\Delta \phi_{\text{m}}(t)-\frac{k_1}{k_1+k_2}\Delta \phi_{\text{m}}(t-T(t))\nonumber.
\end{align}
where $k_1=\omega_{L2}+\omega_2$ and $k_2=\omega_{L1}+\omega_1+\omega_m$.}

\review{For the beat frequency to be detectable, the difference $|\omega_{L1}-\omega_{L2}|$ must be known to within the PLL bandwidth. For a \qty{10}{\kilo\hertz} PLL bandwidth, a \qty{1550}{\nano\metre} optical reference must be known within a fractional frequency of \num{1d-11}. Furthermore, the line-of-sight Doppler shift and velocity must also be known to within the PLL bandwidth, requiring \textit{a priori} knowledge on the order of $\qty{1.55}{\centi\metre\per\second}$.}

By making the first order Taylor expansion $\Delta\phi_m(t-T(t))\approx\Delta\phi_m(t)-T(t)\Delta\dot{\phi}_m(t)$, the TWFT laser difference simplifies to,
\begin{align}
    \Delta\phi_L(t)&=-\frac{k_2}{k_1+k_2}\phi_{\mathrm{1}}(t)+\frac{k_1}{k_1+k_2}\phi_{\mathrm{2}}(t)\label{eq:noiseqn}\\
    &+ \frac{k_2-k_1}{k_1+k_2}\Delta \phi_{\text{m}}(t)-\frac{k_1}{k_1+k_2}T(t)\Delta \dot{\phi}_{\text{m}}(t).\nonumber
\end{align}

\review{This Taylor expansion is sufficient for the experimental demonstration, however, simulations suggest for a LEO link that higher-order terms contribute to stability in parts less than \num{1d-19}, as shown in Fig. S4 in the Supplemental Material \cite{supplemental_material}.}

This frequency difference can then be calculated in post processing with measurements of $\Delta\phi_{1}(t)$, $\Delta\phi_{2}(t)$ and $\Delta\phi_m(t)$. The optical time of flight $T(t)$ is determined from the measured EOM data,
\begin{equation}
    T(t)=T_0 + \frac{\lambda \Delta\phi_m(t)}{c}
\end{equation}
where $T_0$ is an arbitrary time offset, and $\lambda$ is the wavelength of the optical source.

To simplify the experimental demonstration, a folded free-space test link was created using a corner cube retro-reflector (CCR) attached to a drone. This enabled the two optical transceivers to be co-located, reducing the complexity of the signal processing and electronics system design, as shown in Fig.~\ref{fig:ExpDiagram1}(b) and in Fig.~S1 in the Supplemental material \cite{supplemental_material}. 

Each transceiver was provided the same \qty{1550}{\nano\metre} optical source (NKT X15 laser with linewidth \textless \qty{100}{\hertz}; and power of \qty{14}{\deci\belmilliwatt}). A Red Pitaya STEMlab 125-14 software defined radio was used for digitising the PD RF signals, monitoring their phase with all-digital PLLs (used in \cite{McSorley2023}), and coarse servo control of the EOM. \review{This system demonstration is self-referenced, with the Red Pitaya providing a \qty{10}{\mega\hertz} reference to drive $\omega_1$ and $\omega_2$.}

\begin{table}
\centering
\caption{Drone Link Description}
\begin{tabular}{lllll}
\cline{1-3}
Parameter               & Value   &  \\ \cline{1-3}
Link Distance & \SIrange{1.7}{2.3}{\kilo\metre} & \\
T1 Transmit Power exiting Fiber          & \qty{4}{\deci\belmilliwatt}    &  \\
T2 Transmit Power exiting Fiber          & \qty{0}{\deci\belmilliwatt}    &  \\
Typical Link Loss & \qty{30}{\deci\bel}       & \\ 
\review{PLL Bandwidth} & \review{\qty{10}{\kilo\hertz}}       &  \\ 
\review{PD Bandwidth} & \review{\qty{250}{\mega\hertz}}       &  \\ \cline{1-3} 
\end{tabular}
\label{tab:link_budget}

\vspace{-0.5cm}
\end{table}

\newlength{\fthreeh}
\newlength{\fthreew}
\settowidth{\fthreeh}{\includegraphics{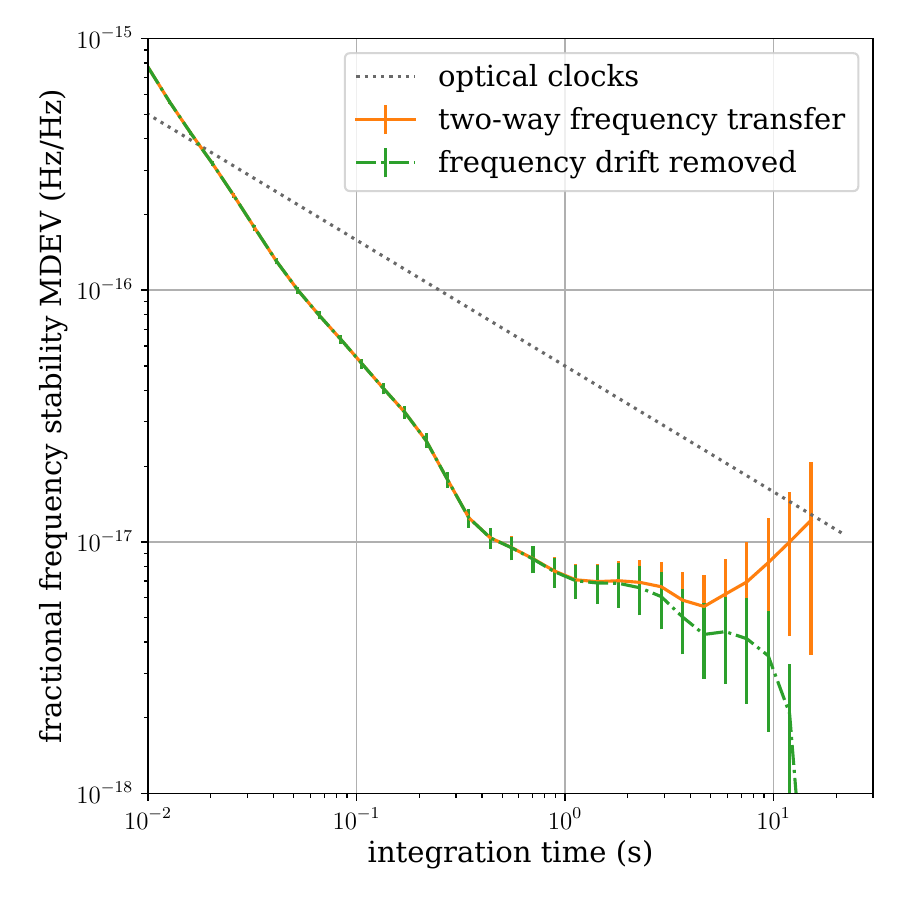}}
\settoheight{\fthreew}{\includegraphics{MDEV.pdf}}
\pgfmathsetmacro{\aspectratiothree}{\fthreew/\fthreeh}
\pgfmathsetmacro{\wordthree}{150 / (\aspectratiothree) + 20}

\newlength{\ffourh}
\newlength{\ffourw}
\settowidth{\ffourh}{\includegraphics{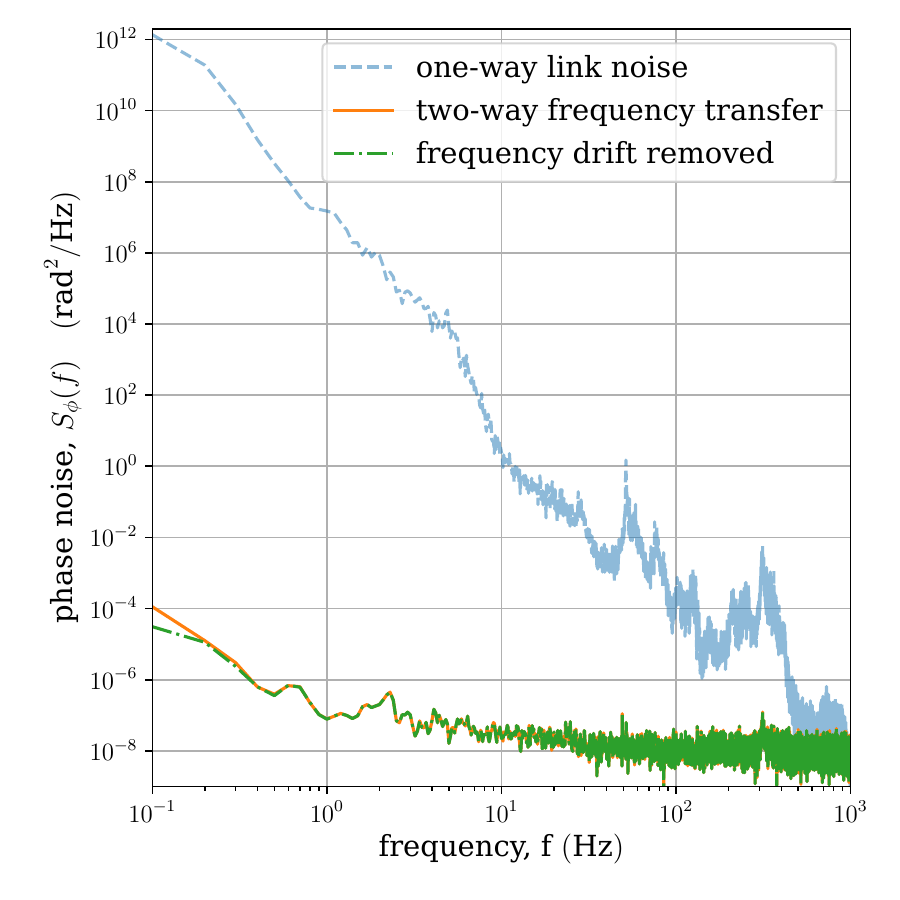}}
\settoheight{\ffourw}{\includegraphics{PSD.pdf}}
\pgfmathsetmacro{\aspectratiofour}{\ffourw/\ffourh}
\pgfmathsetmacro{\wordfour}{150 / (\aspectratiofour) + 20}

\begin{figure*}[ht]
\hspace*{\fill}%
\begin{minipage}[t]{0.45\linewidth}
\centering
\vspace{0pt}
\includegraphics[width=\columnwidth]{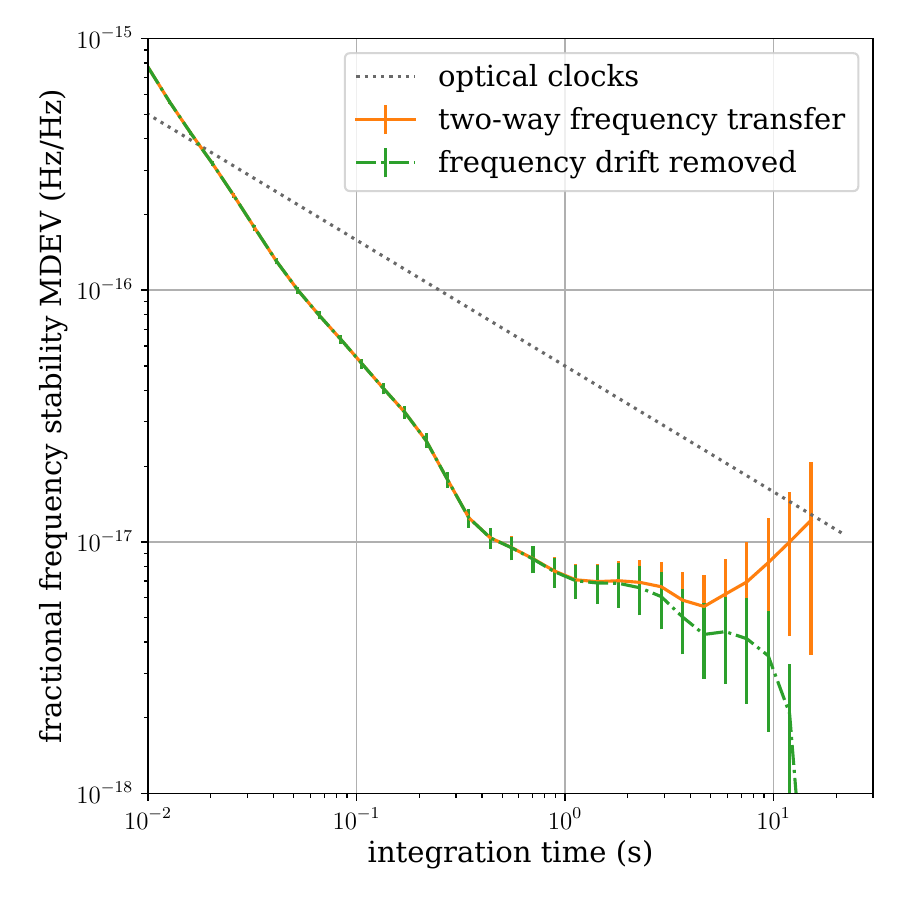}
\caption{Modified Allan deviation for Fig.~\ref{fig:timeseries}(d). Provided are the calculated residual difference (orange), the residual difference with the frequency drift removed (dashed green) and typical state of the art clock performance \cite{Bothwell_2019} (dotted gray).}
\label{fig:MDEV}
\end{minipage}%
\hfill
\begin{minipage}[t]{0.45\linewidth}
\centering
\vspace{0pt}
\includegraphics[width=\columnwidth]{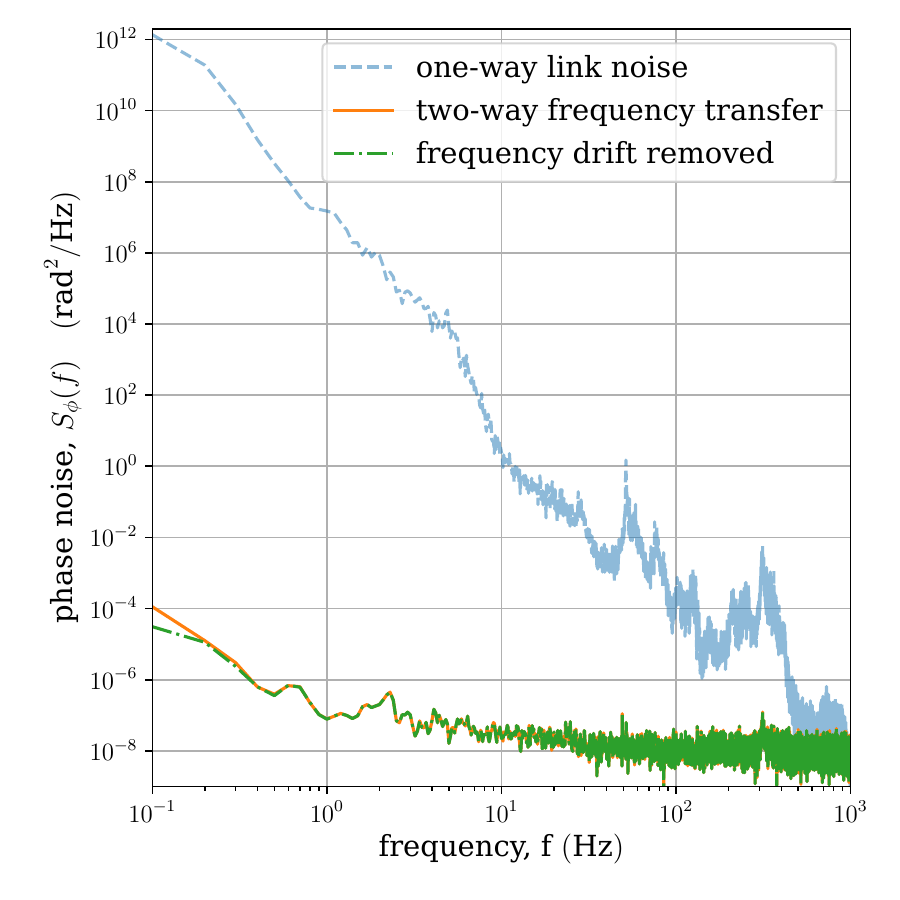}
\caption{Phase power spectral density for Fig.~\ref{fig:timeseries}(d). Provided are the calculated residual difference (orange), the residual difference with the frequency drift removed (dashed green) and the one-way link noise (dashed blue).}
\label{fig:PSD}
\end{minipage}%
\hspace*{\fill}
\end{figure*}

The active terminal described in \cite{Walsh2022} is used to minimise angular pointing errors and maintain link alignment. The nominal link parameters for this moving link are provided in Table \ref{tab:link_budget}. 

A drone manoeuvre involved accelerating the drone at an increasing rate forward and backwards along the line of sight of the optical terminal. The drone was held at maximum speed until loss-of-lock occurred, from either the Doppler tracking or tip-tilt active terminal. Several manoeuvres were undertaken over a \qty{20}{\minute} window.

\review{The PLL bandwidth is dependant on the return power, a result of the mixing stage inside the ADPLL, resulting in loss of lock when large power swings occur. To combat this, two analogue amplifier stages are used prior to digitization that saturate the measured beatnote. A PLL bandwidth of \qty{10}{\kilo\hertz} was chosen to ensure that an initial lock on to PD1 and PD2 was possible, while also minimising the rate of cycle slips. A small bandwidth minimises the impact of shot noise on the PLL \cite{DICK2008, Sambridge2023}, however, the experimental demonstration was primarily limited by the Doppler dynamics and weak optical reflections in the terminal.}
 
A proportional-integral squared controller was designed inside the Red Pitaya to actuate the EOM's frequency $\omega_m(t)$ by the phase error seen on $\phi_{1}(t)$, such that $\omega_m(t)\approx\dot{T}(t)(\omega_L)$. When the controller lost lock, manual re-acquisition of the tracking system was performed, \review{where the drone was stopped and the measurement PLLs were reset.}

The measured PD phases $\Delta\phi_{1}(t)$ and $\Delta\phi_{2}(t)$, and EOM measurement $\Delta\phi_{m}(t)$ were used to calculate the kinematics of the drone for a single manoeuvre. One manoeuvre example is shown in Fig.~\ref{fig:timeseries}. \review{The EOM phase, frequency and frequency rate of change are shown on the left-hand y-axes of Fig.~\ref{fig:timeseries}(a), Fig.~\ref{fig:timeseries}(b) and Fig.~\ref{fig:timeseries}(c). The laser phase and frequency difference, $\Delta\phi_L(t)$ and $\Delta\dot{\phi}_L(t)$, are shown on the left-hand y-axes of Fig.~\ref{fig:timeseries}(d) and Fig.~\ref{fig:timeseries}(e). The range, velocity and acceleration, directly derived from phase and frequency, are also shown on the right-hand y-axes of each plot.}

In this example, the system successfully tracks \qty{300}{\mega\cycle} of accumulated laser phase, or a displacement of \qty{300}{\metre}. \review{The EOM frequency and frequency rate of change show a maximum Doppler shift of \qty{18}{\mega\hertz} at a maximum Doppler rate of \qty{1}{\mega\hertz\per\second}, which corresponds to a maximum line-of-sight velocity \qty{15}{\metre\per\second} at \qty{0.9}{\metre\per\second\squared}.}

\review{To calculate Equation \ref{eq:noiseqn}, an initial time-of-flight, $T_0$, estimate is required. The orange curve in Fig. \ref{fig:timeseries}(d) assumes an initial distance offset of \qty{2106}{\metre}. A parabolic fit of the residual phase indicates a frequency drift of \qty{-200}{\micro\hertz\per\second}. This frequency drift is likely caused by the thermal sensitivity of the quartz oscillator referencing the experiment, as discussed in the Supplemental material \cite{supplemental_material}.}

The laser residual difference, in terms of phase~(range) and frequency~(velocity), is on the order of $\pm$\qty{0.05}{\cycle} ($\pm$\qty{0.1}{\micro\metre}), and $\pm$\qty{100}{\hertz} ($\pm$\qty{100}{\micro\metre\per\second}). 

\review{To minimise the impact of cycle slips on the data set, frequency impulses greater than \qty{800}{Hz} in the residual error are masked as these are assumed to be aphysical and caused by loss-of-lock and cycle slipping. In this example, frequency impulse around $t=\qty{40}{\second}$ are masked, as they caused a significant discontinuity in the residual phase noise.}

The modified Allan deviation (MDEV) fractional frequency stability of the TWFT laser noise, calculated with Eq. \ref{eq:noiseqn}, is provided in Fig.~\ref{fig:MDEV} (solid blue). The MDEV for the residual difference, with the frequency drift removed, is also provided (dashed green). Also provided for comparison is \review{an example of a} state-of-the-art lab-based optical atomic clock (dotted gray) \cite{Bothwell_2019}.

\review{With the frequency drift present, the MDEV integrates down at a slope of roughly $\tau^{-1}$, surpassing optical clocks at $\tau=\qty{0.02}{\second}$, indicating white frequency noise. After, $\tau=\qty{2}{\second}$, the MDEV begins to degrade, attributed to the residual frequency drift. The best obtained MDEV with the frequency drift present is \num{7d-18} at $\tau=\qty{5}{\second}$. By removing the frequency drift, the MDEV improves past $\tau=\qty{5}{\second}$.}

The phase power spectral density (PSD) is also provided in Fig.~\ref{fig:PSD}. The TWFT residual phase noise with the frequency drift (solid blue) and without (dashed green),  and the one-way frequency transfer (dotted blue) are shown for comparison. The one-way PSD is approximated by the $\phi_m(t)$ time-series.

The one-way PSD is dominated by the kinematics of the link, and atmospheric drone noise. Below \qty{1000}{\hertz}, the two-way PSD shows improvement on the order of \qty{d13}{\radian\squared\per\hertz} at \qty{1}{\hertz}, and \qty{d4}{\radian\squared\per\hertz} at \qty{100}{\hertz}. Between \qty{1}{\hertz} and \qty{1000}{\hertz}, this compensated PSD noise follows a $f^0$ profile, with a residual noise floor of \qty{d-8}{\radian\squared\per\hertz}, indicative of white phase noise. This white noise floor is likely a result of the analogue amplification chain.

Between \qty{1}{\hertz} and \qty{10}{\hertz}, the compensated PSD follows a $f^{-1}$ profile, which becomes worse below \qty{1}{\hertz}, following a $f^{-3}$ profile. \review{This low frequency increase in phase noise can again be attributed to the frequency drift in the residual phase. The PSD of the residual phase, with the drift removed, shows slight improvement below \qty{0.1}{\hertz}, however it still follows a $f^{-3}$ profile. It is possible that imperfect Doppler compensation may contribute to this increase in noise.}

In summary, we have demonstrated optical TWFT over a rapidly moving link.  This system can be extended to ground-to-space optical frequency transfer with the addition of an \textit{a priori} EOM frequency sweep. This \textit{a priori} sweep can be determined for satellites in LEO \cite{Chiodo2013}, with frequency sweeps capable of supporting Doppler shifts \textgreater \qty{20}{\giga\hertz} and Doppler rates \textgreater \qty{100}{\mega\hertz}.

\section*{Acknowledgments}
This work has been supported by the SmartSat CRC, whose activities are funded by the Australian Government’s CRC Program. S.M.P.M, A.S.M and A.M.F are supported by Australian Government Research Training Program Scholarships and top-up scholarships funded by the Government of Western Australia. S.K is supported by a SmartSat CRC PhD Scholarship and a research award from The Andy Thomas Space Foundation and EOS Space Systems. \\

\input{apssamp.bbl}
\cleardoublepage
\stoprecording

\ifnum\value{highlights}>0
\section*{List of Modifications}
\highlightslist
\else
\fi

\end{document}

%% file: fig1.pdf_tex
\begingroup%
  \makeatletter%
  \providecommand\color[2][]{%
    \errmessage{(Inkscape) Color is used for the text in Inkscape, but the package 'color.sty' is not loaded}%
    \renewcommand\color[2][]{}%
  }%
  \providecommand\transparent[1]{%
    \errmessage{(Inkscape) Transparency is used (non-zero) for the text in Inkscape, but the package 'transparent.sty' is not loaded}%
    \renewcommand\transparent[1]{}%
  }%
  \providecommand\rotatebox[2]{#2}%
  \newcommand*\fsize{\dimexpr\f@size pt\relax}%
  \newcommand*\lineheight[1]{\fontsize{\fsize}{#1\fsize}\selectfont}%
  \ifx\svgwidth\undefined%
    \setlength{\unitlength}{507.96849529bp}%
    \ifx\svgscale\undefined%
      \relax%
    \else%
      \setlength{\unitlength}{\unitlength * \real{\svgscale}}%
    \fi%
  \else%
    \setlength{\unitlength}{\svgwidth}%
  \fi%
  \global\let\svgwidth\undefined%
  \global\let\svgscale\undefined%
  \makeatother%
  \begin{picture}(1,0.36149554)%
    \lineheight{1}%
    \setlength\tabcolsep{0pt}%
    \put(0,0){\includegraphics[width=\unitlength,page=1]{fig1.pdf}}%
    \put(0.93558631,0.02868079){\color[rgb]{0,0,0}\makebox(0,0)[lt]{\lineheight{0.69999999}\smash{\begin{tabular}[t]{l}\textbf{CCR}\end{tabular}}}}%
    \put(0,0){\includegraphics[width=\unitlength,page=2]{fig1.pdf}}%
    \put(0.91097597,0.10144661){\makebox(0,0)[t]{\lineheight{0.69999999}\smash{\begin{tabular}[t]{c}\textbf{Drone}\end{tabular}}}}%
    \put(0,0){\includegraphics[width=\unitlength,page=3]{fig1.pdf}}%
    \put(0.90458054,0.03015727){\color[rgb]{0,0,0}\makebox(0,0)[t]{\lineheight{1.12}\smash{\begin{tabular}[t]{c}$T(t)$\end{tabular}}}}%
    \put(0,0){\includegraphics[width=\unitlength,page=4]{fig1.pdf}}%
    \put(0.63815329,0.09825986){\makebox(0,0)[lt]{\lineheight{1.12}\smash{\begin{tabular}[t]{l}\textbf{Transceiver 1}\end{tabular}}}}%
    \put(0.63817015,0.05549416){\makebox(0,0)[lt]{\lineheight{1.12}\smash{\begin{tabular}[t]{l}\textbf{Transceiver 2}\end{tabular}}}}%
    \put(0.50783471,0.136786){\makebox(0,0)[lt]{\lineheight{1.12}\smash{\begin{tabular}[t]{l}\textbf{Experimental Demonstration}\end{tabular}}}}%
    \put(0.51036314,0.04639841){\color[rgb]{0,0,0}\makebox(0,0)[lt]{\lineheight{1.25}\smash{\begin{tabular}[t]{l}Optical\end{tabular}}}}%
    \put(0.51036314,0.02720435){\color[rgb]{0,0,0}\makebox(0,0)[lt]{\lineheight{1.25}\smash{\begin{tabular}[t]{l}Source\end{tabular}}}}%
    \put(0.50150436,0.06559255){\color[rgb]{0,0,0}\makebox(0,0)[lt]{\lineheight{1.25}\smash{\begin{tabular}[t]{l}Common\end{tabular}}}}%
    \put(0,0){\includegraphics[width=\unitlength,page=5]{fig1.pdf}}%
    \put(0.95978279,0.3409875){\color[rgb]{0,0,0}\makebox(0,0)[lt]{\lineheight{1.25}\smash{\begin{tabular}[t]{l}\textbf{a)}\end{tabular}}}}%
    \put(0.97259004,0.1388767){\color[rgb]{0,0,0}\makebox(0,0)[lt]{\lineheight{1.25}\smash{\begin{tabular}[t]{l}\textbf{b)}\end{tabular}}}}%
    \put(0,0){\includegraphics[width=\unitlength,page=6]{fig1.pdf}}%
    \put(0.24509395,0.15331328){\makebox(0,0)[t]{\lineheight{1.12}\smash{\begin{tabular}[t]{c}PD 1\end{tabular}}}}%
    \put(0,0){\includegraphics[width=\unitlength,page=7]{fig1.pdf}}%
    \put(0.81758432,0.21333472){\rotatebox{90}{\makebox(0,0)[t]{\lineheight{1.12}\smash{\begin{tabular}[t]{c}PD 2\end{tabular}}}}}%
    \put(0,0){\includegraphics[width=\unitlength,page=8]{fig1.pdf}}%
    \put(0.34651782,0.2660024){\makebox(0,0)[t]{\lineheight{1.12}\smash{\begin{tabular}[t]{c}AOM 1\end{tabular}}}}%
    \put(0,0){\includegraphics[width=\unitlength,page=9]{fig1.pdf}}%
    \put(0.14276466,0.2660024){\makebox(0,0)[t]{\lineheight{1.12}\smash{\begin{tabular}[t]{c}EOM\end{tabular}}}}%
    \put(0,0){\includegraphics[width=\unitlength,page=10]{fig1.pdf}}%
    \put(0.75740024,0.26600561){\makebox(0,0)[t]{\lineheight{1.12}\smash{\begin{tabular}[t]{c}AOM 2\end{tabular}}}}%
    \put(0,0){\includegraphics[width=\unitlength,page=11]{fig1.pdf}}%
    \put(0.24562726,0.08137978){\makebox(0,0)[t]{\lineheight{1.12}\smash{\begin{tabular}[t]{c}BPF\end{tabular}}}}%
    \put(0,0){\includegraphics[width=\unitlength,page=12]{fig1.pdf}}%
    \put(0.14026461,0.08096626){\makebox(0,0)[t]{\lineheight{1.12}\smash{\begin{tabular}[t]{c}CFT\end{tabular}}}}%
    \put(0.09237022,0.01781089){\makebox(0,0)[t]{\lineheight{1.12}\smash{\begin{tabular}[t]{c}Optical Path\end{tabular}}}}%
    \put(0.2456003,0.01781089){\makebox(0,0)[t]{\lineheight{1.12}\smash{\begin{tabular}[t]{c}Electrical Path\end{tabular}}}}%
    \put(0.08742065,0.31136847){\makebox(0,0)[lt]{\lineheight{1.12}\smash{\begin{tabular}[t]{l}\textbf{Transceiver 1}\end{tabular}}}}%
    \put(0.09079499,0.23921591){\color[rgb]{0,0,0}\makebox(0,0)[lt]{\lineheight{1.12}\smash{\begin{tabular}[t]{l}$\omega_{m(t)}$\end{tabular}}}}%
    \put(0.35567548,0.30702788){\color[rgb]{0,0,0}\makebox(0,0)[lt]{\lineheight{1.12}\smash{\begin{tabular}[t]{l}$\omega_1$\end{tabular}}}}%
    \put(0.72292763,0.30702757){\color[rgb]{0,0,0}\makebox(0,0)[lt]{\lineheight{1.12}\smash{\begin{tabular}[t]{l}$\omega_2$\end{tabular}}}}%
    \put(0.02231414,0.30894553){\color[rgb]{0,0,0}\makebox(0,0)[lt]{\lineheight{1.12}\smash{\begin{tabular}[t]{l}$\phi_{L1}(t)$\end{tabular}}}}%
    \put(0.79235725,0.31338704){\makebox(0,0)[lt]{\lineheight{1.12}\smash{\begin{tabular}[t]{l}\textbf{Transceiver 2}\end{tabular}}}}%
    \put(0.93494323,0.30894553){\color[rgb]{0,0,0}\makebox(0,0)[lt]{\lineheight{1.12}\smash{\begin{tabular}[t]{l}$\phi_{L2}(t)$\end{tabular}}}}%
    \put(0.7085839,0.20721354){\color[rgb]{0,0,0}\makebox(0,0)[lt]{\lineheight{1.12}\smash{\begin{tabular}[t]{l}$V_{2}(t)$\end{tabular}}}}%
    \put(0.25099983,0.117878){\color[rgb]{0,0,0}\makebox(0,0)[lt]{\lineheight{1.12}\smash{\begin{tabular}[t]{l}$V_{1}(t)$\end{tabular}}}}%
    \put(0.01735442,0.34075457){\makebox(0,0)[lt]{\lineheight{1.12}\smash{\begin{tabular}[t]{l}\textbf{Point to Point Architecture}\end{tabular}}}}%
    \put(0,0){\includegraphics[width=\unitlength,page=13]{fig1.pdf}}%
    \put(0.00966198,0.23020636){\color[rgb]{0,0,0}\makebox(0,0)[lt]{\lineheight{1.25}\smash{\begin{tabular}[t]{l}Optical\end{tabular}}}}%
    \put(0.01114183,0.21267473){\color[rgb]{0,0,0}\makebox(0,0)[lt]{\lineheight{1.25}\smash{\begin{tabular}[t]{l}Source 1\end{tabular}}}}%
    \put(0.92722886,0.22843528){\color[rgb]{0,0,0}\makebox(0,0)[lt]{\lineheight{1.25}\smash{\begin{tabular}[t]{l}Optical\end{tabular}}}}%
    \put(0.92575586,0.21090365){\color[rgb]{0,0,0}\makebox(0,0)[lt]{\lineheight{1.25}\smash{\begin{tabular}[t]{l}Source 2\end{tabular}}}}%
    \put(0.55498992,0.23117059){\color[rgb]{0,0,0}\makebox(0,0)[t]{\lineheight{0}\smash{\begin{tabular}[t]{c}$T(t)$\end{tabular}}}}%
    \put(0,0){\includegraphics[width=\unitlength,page=14]{fig1.pdf}}%
  \end{picture}%
\endgroup%

%% file: apssamp.bbl
%